\documentclass[letter,draftcls,onecolumn,10pt,twoside]{IEEEtranTCOM}
\usepackage{units}
\usepackage{graphicx,epsfig,xcolor}
\usepackage{amssymb}
\usepackage{amscd}
\usepackage{eucal}
\usepackage[cmex10]{amsmath}
\usepackage{cite,soul}
\begin{document}
\pagestyle{empty}
\title{{Comparison of OFDM and SC-DFE Capacities \\Without Channel Knowledge at the Transmitter}}

\author{Amanda de Paula and Cristiano Panazio
\thanks{Cristiano Panazio and Amanda de Paula are with
Departamento de Engenharia de Telecomunica\c{c}\~{o}es e Controle (PTC), Escola Polit\'{e}cnica, University of S\~{a}o Paulo, SP, Brazil, e-mails: \{cpanazio,amanda\}@lcs.poli.usp.br}%
 }

\maketitle

\begin{abstract}
This letter provides a capacity analysis between OFDM and the ideal SC-DFE when no channel knowledge is available at the transmitter. Through some algebraic manipulation of the OFDM and SC-DFE capacities and using the concavity property of the manipulated capacity function and Jensen's inequality, we are able to prove that the SC-DFE capacity is always superior to that of an OFDM scheme for 4- and 16-QAM for any given channel. For higher-order modulations, however, the results indicate that OFDM may only surpass the ideal SC-DFE capacity by a small amount in some specific scenarios.


\end{abstract}


\section{Introduction}


\IEEEPARstart{I}{t} is known that orthogonal frequency division multiplexing (OFDM) without channel knowledge at the transmitter, such as the case in digital radio and television broadcasting, depends on channel coding to achieve good performance \cite{Sari95,Wilson95,Wang04}.
Channel coding allows OFDM to exploit the channel frequency diversity by using the non-faded subcarriers to recover the information carried on attenuated subcarriers. By contrast, the single-carrier (SC) scheme is able to exploit such diversity even in the absence of channel coding, since each transmitted symbol spreads throughout the entire used band due to its smaller duration when compared to OFDM. Furthermore, the cyclic-prefix (CP) and the one-tap equalizer techniques, which allow low complexity equalization, are not a privilege of OFDM and can as well be applied to the SC scheme \cite{Falconer02}. It is also possible to enhance the performance of the SC scheme with little additional complexity by using the decision-feedback equalizer (DFE). In addition, it was shown by \cite{Cioffi95_p1} that using a DFE without error propagation, \emph{i.e.}, an ideal DFE, with an unbiased minimum mean square error (MMSE) criterion the channel capacity can be achieved. These facts have spawned many comparisons between OFDM and SC.

Analytical results were provided by \cite{Wilson95} and \cite{Aue98}. The latter, by using the cutoff rate, analyzes the effect of the coding rate considering 4-quadrature amplitude modulation (QAM), a two-tap block-fading channel configuration scenario and just a SC with linear equalization. The former shows that both the OFDM and SC-DFE schemes can exploit the frequency diversity on frequency selective block-fading channels, but there are no results on channel capacities differences. This subject is addressed in \cite{Franceschini08}, in which the authors conjecture that the SC scheme with an optimal receiver has equal or better performance than OFDM for a uniform power spectrum transmission. It also considers that the possible rate difference will be larger when the channel is more frequency selective. 

In this letter, we analyze such behavior using channel capacity results and Jensen's inequality \cite{Cover06} and we show that the conjecture presented by \cite{Franceschini08} is valid for any given channel but only for 4- and 16-QAM. For higher-order modulations, the conjecture is true in almost all scenarios, but in some very particular cases the OFDM scheme can present some marginal capacity advantage over the ideal SC-DFE system. 

This letter is organized as follows. In Section \ref{sec:System_Model}, we describe the system model. In Section \ref{sec:capacity}, the comparison between the schemes is done through the use of the channel capacity for different modulation cardinalities and any channel, using concavity function analysis and Jensen's inequality. Finally, conclusions are stated in Section \ref{sec:conclusions}.

\section{System Model}
\label{sec:System_Model}

In block transmission schemes using the CP approach, before transmitting each block $\mathbf{s}=\left[s_0,\cdots, s_{N-1}\right]^T$, it is appended a CP of length $L-1$. Theses samples are linearly convoluted with a channel $\mathbf{h}=\left[h_0,\cdots, h_{L-1},0,\cdots,0\right]^T$, with $L<N$ and a zero-mean circularly complex white Gaussian noise with variance $\sigma_v^2$ is added to the channel output. By removing the CP at the receiver, the received signal can be represented by $\mathbf{r}=\boldsymbol{\mathcal{H}}\mathbf{s}+\mathbf{v}$, where $\boldsymbol{\mathcal{H}}$ is a circulant matrix whose first column is given by $\mathbf{h}$ and $\mathbf{v}=\left[v_0,\cdots, v_{N-1}\right]^T$ is the added noise. Without loss of generality, we assume that $\mathbf{h}$ has unitary norm. In the following, we build the OFDM and SC-DFE schemes upon this model.

\subsection{Orthogonal Frequency Division Multiplexing}
Consider $\mathbf{x}=\left[x_0,\cdots, x_{N-1}\right]^T$ the block of QAM symbols to be transmitted in $N$ subcarriers. This is achieved by multiplying $\mathbf{x}$ by the unitary inverse discrete Fourier transform (IDFT) matrix of dimension $N\times N$, which results in $\mathbf{s}=\mathbf{F}^H\mathbf{x}$. In the receiver, we apply a discrete Fourier transform (DFT) $\mathbf{F}$ to $\mathbf{r}$ and we obtain $\mathbf{y}=\mathbf{H}\mathbf{x}+\mathbf{Fv}$,
where $\mathbf{H}=\mathbf{F}\boldsymbol{\mathcal{H}}\mathbf{F}^H=\rm{diag}\{\mathbf{Fh}\}$, and $\mathbf{Fv}$ is also a zero-mean circularly complex white Gaussian noise with variance $\sigma_v^2$, due to the use of the unitary DFT.
We can interpret each element of $\mathbf{y}$ as an additive white Gaussian noise (AWGN) channel with a complex gain. Thus, the equalizer used to estimate the transmitted symbols can be reduced to phase and magnitude compensations, the so-called one-tap equalizer, which do not change the SNR in each subcarrier. Then, the estimated symbols are $\tilde{\mathbf{x}}=\mathbf{Q}\mathbf{y},$ where ${\mathbf{Q}}=\rm{diag}\{\left[Q_0,\cdots, Q_{N-1}\right]\}=\mathbf{H}^{-1}$. 

\subsection{Single Carrier-Decision Feedback Equalizer}
In order to make the capacity comparison feasible, we have to make some simplifications. The first one is that there is no error propagation, since it is hard to analyze and it is known to considerably affect the DFE performance. In practice, this can approximated by using iterative block DFE schemes that can account for the reliability of the decisions that are fed back at the expense of some additional computational complexity \cite{Ben05, Sari06, Hanzo11}. In this paper, in order to eliminate the error propagation, we assume that the transmitted sequence is known at the reception and is fed back into the feedback filter. The second assumption is related to the causality imposed by the feedback filter and the constraint imposed by the circular convolution needed for the one-tap equalization. This requires to initialize the feedback filter inputs with the CP samples, but these symbols will only be decided by the end of the block, which implies in non-casuality. In practice, this can be solved by using the unique word (UW) technique \cite{Witschnig02,Falconer02} that is a fixed pseudo-random sequence known to the receiver and that is used in place of the CP. Nevertheless, if we use the same DFT length as the one used with the CP technique, the UW technique has a lower spectral efficiency. A possible solution to make both modulation schemes have almost the same efficiency is to consider the DFT block much larger than the UW or CP lengths. However, in order to simplify calculations and to keep the same transmission structure, we assume the non-casuality hypothesis and use the CP method in the analysis.

The SC scheme can be viewed as a linearly precoded OFDM where the symbols $\mathbf{x}$ are precoded by the unitary DFT matrix $\mathbf{F}$ (\emph{i.e.}, $\mathbf{s}=\mathbf{F}^H\mathbf{F}\mathbf{x}=\mathbf{x}$) and the received signal, after the one-tap equalizer, must be decoded by the unitary IDFT matrix $\mathbf{F}^H$. 
However, SC allows us to improve its performance by using a non-linear filter, the DFE, which uses past decisions on the received signal to remove intersymbol-interference. Thus, using the hypotheses discussed in the beginning of this subsection, the SC-DFE output can be written as $\tilde{\mathbf{x}}=\mathbf{F}^H\left[\mathbf{QF}\boldsymbol{\mathcal{H}}+\mathbf{BF}\right]\mathbf{x}+\mathbf{F}^H\mathbf{QFv}$, where ${\mathbf{Q}}=\rm{diag}\{\left[Q_0,\cdots, Q_{N-1}\right]\}$ and ${\mathbf{B}}=\rm{diag}\{\left[B_0,\cdots, B_{N-1}\right]\}$ are the feedforward and feedback filters in the frequency domain respectively.

The chosen criterion to obtain the filter coefficients is the unbiased MMSE criterion, since its is shown by \cite{Cioffi95_p1} that an ideal DFE with such criterion can achieve channel's capacity. The coefficients can be calculated as shown in \cite{Benvenuto02,Falconer_DFE02}. Also, with regard to the MMSE SC-DFE, the number of feedback coefficients that maximizes the performance is equal to the channel's memory \cite{Valcarce04}.

\section{Capacity Analysis}
\label{sec:capacity}
In this section, we analyze the capacity difference between the ideal SC-DFE and OFDM schemes. Initially, in order to calculate the OFDM capacity, let us define the signal-to-noise ratio (SNR) for each subcarrier as:
\begin{equation}
\gamma_k=\gamma\left|H_k\right|^2,
\end{equation}
where $H_k$ is the $k^{th}$ element of the diagonal of $\mathbf{H}$ and $\gamma=\frac{\sigma_x^{2}}{\sigma_v^2}$ is the average SNR, since $\sum_{k=0}^{N-1}\left|H_k\right|^2=1$, due to the unitary norm channel and the unitary DFT.

Let us first consider a Gaussian input. In such a case, the capacity per real dimension for a uniform power spectrum allocation input is the average capacity of the subcarriers:
\begin{equation}
C_{\mathrm{OFDM}}=\frac{1}{N}\sum_{k=0}^{N-1}{\frac{1}{2}\log_2\left(1+\gamma_k\right)}.
\label{eq:cap_ofdm_gauss}
\end{equation}

It is noteworthy that a single channel code can be used to code the information bits to form $\mathbf{x}$ as long as the information rate is lower than \eqref{eq:cap_ofdm_gauss} \cite{Root68}.

For the ideal SC-DFE scheme, in order to calculate its capacity, we must obtain the SNR at its output. Let us first assume a system without CP and an infinite length feedforward filter. In such a case, it is known that the SNR at the output of such MMSE DFE equalizer is given by \cite[p.663]{Proakis08}:
\begin{equation}
\gamma_{\mathrm{DFE},\infty}=\exp\left\{\frac{T}{2\pi}\int_{-\pi/T}^{\pi/T}\log\left(\frac{\sigma_v^2+\sigma_x^2\left|H(e^{j\omega T})\right|^2}{\sigma_v^2}\right)d\omega\right\}-1,
\label{eq:SNR_dfe_gauss_infinity}
\end{equation}
where $H(e^{j\omega T})$ is the discrete time Fourier transform of the channel impulse response at the frequency $\omega$ and $T$ is the symbol period.

Since the CP guarantees that we can perfectly invert a FIR channel that does not have spectral nulls with another FIR filter and that the transmission power and bandwidth are kept the same when the CP is added, in order to obtain the SNR at the SC-DFE output we just have to discretize the SNR given by (\ref{eq:SNR_dfe_gauss_infinity}) at the frequencies $\omega(k)=2\pi k/NT$:
\begin{equation}
\gamma_\mathrm{DFE}=\exp\left\{\frac{1}{N}\sum_{k=0}^{N-1}{\log\left(1+\gamma_k\right)}\right\}-1,
\label{eq:DFE_SNR}
\end{equation}
since $\frac{\sigma_v^2+\sigma_x^2\left|H\left(e^{j\omega(k)T}\right)\right|^2}{\sigma_v^2}=1+\gamma_k$.

Therefore, also assuming that a Gaussian modulation is used, so that the equalizer output is also Gaussian, the capacity per real dimension of the ideal SC-DFE is:
\begin{equation}
C_{\mathrm{DFE}}=\frac{1}{2}\log_2\left(1+\gamma_\mathrm{DFE}\right).
\label{eq:cap_dfe_gauss}
\end{equation}

By placing \eqref{eq:DFE_SNR} in \eqref{eq:cap_dfe_gauss}, it results exactly in \eqref{eq:cap_ofdm_gauss}; hence, the OFDM and ideal SC-DFE systems present the same capacity.

However, such analysis consider that Gaussian symbols are transmitted, which is not the case in practice. For ordinary modulations, such as $M$-ary QAM, the capacity can still be numerically evaluated. Considering an AWGN channel and only square $M$-QAM schemes, the associated capacity is two times the capacity of a $\sqrt{M}$-pulse amplitude modulation \cite{Cover06}:
\begin{equation}
C^{\text{M-QAM}}(\gamma)=-2\int\limits_{-\infty}^{+\infty}{f_{R}\left(r,\gamma\right)\log_2f_{R}\left(r,\gamma\right)dr}-\log_2{\frac{2\pi e\sigma_x^2}{\gamma}},
\label{eq:Cap_mqam}
\end{equation}
where $f_{R}\left(r,\gamma\right)=\sqrt{\frac{\gamma}{2M\pi\sigma_x^2}}\sum_{m=-\sqrt{M}/2}^{\sqrt{M}/2-1}e^{-\frac{\gamma\left(r-(2m+1)\right)^2}{2\sigma_x^2}},$
which corresponds to the sum of Gaussian distributions with variance $\sigma_v^2=\gamma/\sigma_x^2$, centered at the $\sqrt{M}$-PAM points and pondered by $1/\sqrt{M}$ in order to have unitary area.

Equation (\ref{eq:Cap_mqam}) has a shape similar to the Gaussian capacity, but it has an asymptotic gap of 1.53 dB (\emph{i.e.}, the shaping gain) when $M\to\infty$ and it will saturate at $\log_2M$ for finite $M$. This saturation is the main reason that OFDM in frequency selective channels will present a capacity degradation when compared to the ideal SC-DFE scheme. Due to the saturation, for certain average SNR values, some subcarriers cannot attain a capacity close to what would be attained by a larger $M$ scenario for the same average SNR values and are not able to fulfill the expected capacity to compensate for the attenuated subcarriers. To the best of the authors knowledge, such limitation has been only observed in references \cite{Franceschini08} and \cite{Wesel95}, but the latter misses a detailed explanation and, in the former, although it provides compelling evidence, it does not provide conclusive proofs on the capacity comparison of OFDM with SC schemes.

Before establishing a detailed analysis, let us first illustrate the origin of the OFDM capacity limitations. In order to show this, let us consider that any residual ISI of the QAM signal in the ideal SC-DFE output can be modeled as a Gaussian noise due to the central limit theorem. Hence, the capacity of the ideal SC-DFE system can be evaluated applying \eqref{eq:DFE_SNR} in~\eqref{eq:Cap_mqam}:
\begin{equation}
C^{\text{M-QAM}}_{\text{DFE}}=C^{\text{M-QAM}}\left(\gamma_{\mathrm{DFE}}\right).
\label{eq:Cap_mqam_sccp}
\end{equation}

On the other hand, the OFDM capacity is the average of the capacities in the different subcarriers:
\begin{equation}
\label{eq:Cap_mqam_ofdm}
C^{\text{M-QAM}}_{\text{OFDM}}=\frac{1}{N}\sum_{k=0}^{N-1}{C^{\text{M-QAM}}\left(\gamma_k\right)}.
\end{equation}

If we have $\gamma\rightarrow\infty$, then (\ref{eq:Cap_mqam_sccp}) and (\ref{eq:Cap_mqam_ofdm}) converge to $\log_2M$. If $\gamma$ is small enough for all $k$ to have $C^{\text{M-QAM}}(\gamma_k)\approx C^{\text{M}'\text{-QAM}}(\gamma_k)$, with {$M'\gg~M$}, then the capacities are practically equal. However, if $C^{\text{M-QAM}}(\gamma_k)$ falls close to the saturation region of (\ref{eq:Cap_mqam}) for certain values of $k$, then the OFDM capacity will be inferior to that of the ideal SC-DFE.

\begin{figure}
\centering
\includegraphics[scale=.47]{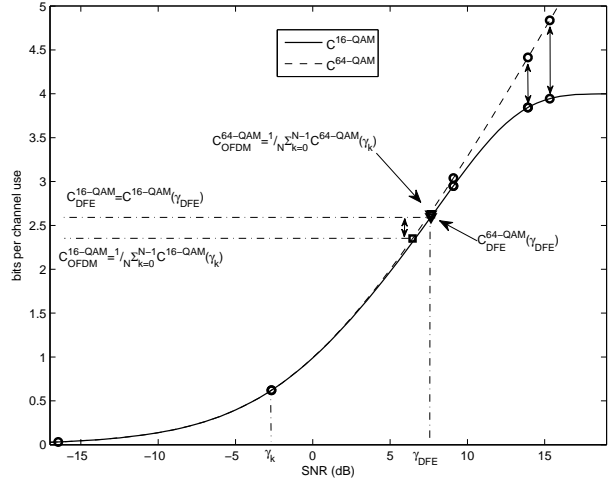}\vspace{-0.2cm}
\caption{Capacity for the OFDM and SC-DFE systems considering 16-QAM and 64-QAM for a SNR value equal to $\gamma=11$ dB. The ($\nabla$) and the ($\Box$) represent the SC-DFE and OFDM capacities respectively and the (o) represents the capacity of each subcarrier.}
\label{Fig:cap_highsnr}
\end{figure}

As an example and without loss of generality, we calculate the OFDM and SC-DFE capacities for the unitary norm channel with zeros in $0.95\exp\left(\pm j 0.9\pi\right)$, considering 16- and 64-QAM modulations, $N=8$ and $\gamma= 11$ dB. The results are depicted in Fig.~\ref{Fig:cap_highsnr}, where we show the ideal SC-DFE and OFDM capacities, as well as the OFDM capacity of each one of its subcarriers. 
As it can be seen, the OFDM and SC-DFE capacities are practically the same for 64-QAM. However, for the same average SNR, when using 16-QAM, there is a capacity difference between them. This is due to the fact that the capacity of some subcarriers (the ones with higher gains) falls close to the capacity saturation region and cannot compensate the lower capacity of the attenuated subcarriers. It is worth noting that the capacity of the SC-DFE has barely changed, since, for such $\gamma_{DFE}$, which does not depends on the cardinality, the capacity of both 16- and 64-QAM are almost the same.

From the results discussed in the previous paragraph, we can predict that larger deviations of $\left|H_k\right|^2$ will be more prone to create differences between the capacities of the OFDM and ideal SC-DFE schemes. 

In the next subsection, we show 
that the ideal SC-DFE is always better than OFDM for 4- and 16-QAM, and that for higher cardinalities, OFDM can be marginally better in some specific cases.

\subsection{Theoretical Capacity Comparison between the ideal SC-DFE and OFDM using QAM}\label{ap:Cap_teo}

In order to compare \eqref{eq:Cap_mqam_sccp} and \eqref{eq:Cap_mqam_ofdm}, firstly, let us define $\phi\left(x\right)=\log\left(x+1\right)$ and, thus, $\gamma_{DFE}$ can be expressed as:
\begin{equation}
\gamma_{DFE}=\phi^{-1}\left(\frac{1}{N}\sum_{k=0}^{N-1}\phi\left(\gamma_k\right)\right).
\end{equation}
In the following, we can write:
\begin{equation}
\tau\left(x\right)=C^{\text{M-QAM}}\left(\phi^{-1}(x)\right),
\end{equation}
so that \eqref{eq:Cap_mqam_sccp} and \eqref{eq:Cap_mqam_ofdm} can be rewritten as
\begin{equation}
C^{\text{M-QAM}}_{\text{DFE}}\left(\gamma\right)=\tau\left(\frac{1}{N}\sum_{k=0}^{N-1}\phi\left(\gamma_k\right)\right),
\end{equation}
and
\begin{equation}
C^{\text{M-QAM}}_{\text{OFDM}}\left(\gamma\right)=\frac{1}{N}\sum_{k=0}^{N-1}\tau\left(\phi\left(\gamma_k\right)\right).
\end{equation}

Therefore, all we have to do is to analyze the concavity properties of:
\begin{equation}
\label{eq:tau}
\tau\left(x\right)=C^{\text{M-QAM}}\left(\exp(x)-1\right)
\end{equation}

The concavity properties of \eqref{eq:tau} can be analyzed through its second derivative. If it is positive, we can state that the function $\tau\left(x\right)$ is convex in the considered interval. Otherwise, we state that $\tau\left(x\right)$ is concave. In this case, using Jensen's inequality, we state that:
\begin{equation}
\tau\left(\frac{1}{N}\sum_{k=0}^{N-1}\phi\left(\gamma_k\right)\right) \geq \frac{1}{N}\sum_{k=0}^{N-1}\tau\left(\phi\left(\gamma_k\right)\right),
\end{equation}
which means that $C^{\text{M-QAM}}_{\text{DFE}}\left(\gamma\right)\geq C^{\text{M-QAM}}_{\text{OFDM}}\left(\gamma\right)$.

\begin{figure}[!t]
\centering
\includegraphics[scale=.5]{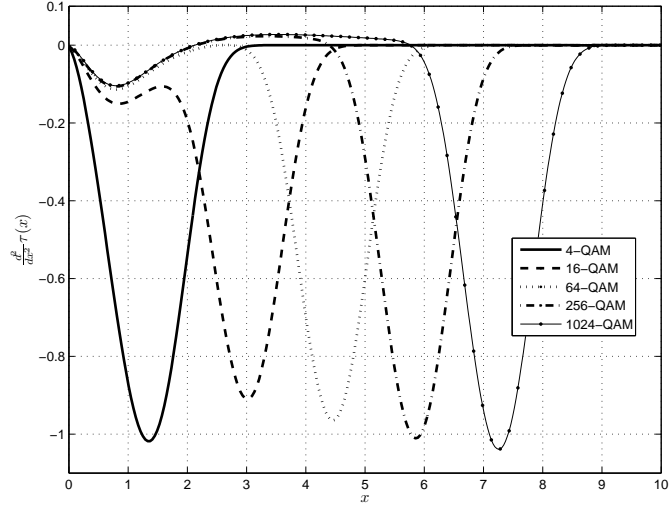}\vspace{-0.2cm}
\caption{Second derivative of $\tau(x)$}
\label{Fig:concavity}
\end{figure}

\begin{figure}[h]
\centering
\includegraphics[scale=.5]{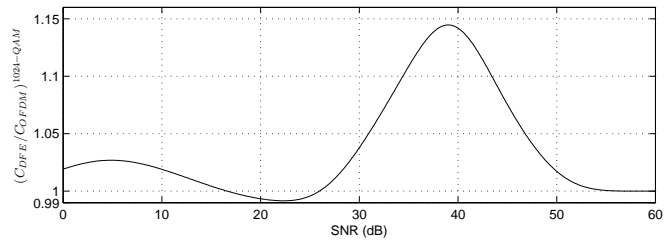}\vspace{-0.2cm}
\caption{Capacity ratio between $C^{1024\text{-QAM}}_{\mathrm{DFE}}$ and $C^{1024\text{-QAM}}_{\mathrm{OFDM}}$ for 1024-QAM, $H(z)=0.1624(1+z^{-4})+0.4546(z^{-1}+z^{-3})+0.7307z^{-2}$ and N=512.}
\label{Fig:comp_cap}
\end{figure}

In Fig. \ref{Fig:concavity}, we show the second derivative of the function $\tau(x)$, numerically evaluated, for different values of $M$. We can observe that it is a non-positive function for $M$ equal to 4 and 16. For the other cardinalities, there are some intervals of $x$ that the second derivative is positive and thus OFDM may present a higher capacity than the ideal SC-DFE depending on the channel. Anyway, the second derivative of $\tau(x)$ assumes only small positive values from which we conclude that the performance advantage tends to be small, even for larger cardinalities. Such behavior is also observed for larger $M$, where such interval of $x$ gets broader. For instance, the ratio $C^{\text{M-QAM}}_{\text{DFE}}/C^{\text{M-QAM}}_{\text{OFDM}}$ for the channel $H(z)=0.1624(1+z^{-4})+0.4546(z^{-1}+z^{-3})+0.7307z^{-2}$ is shown in Fig.~\ref{Fig:comp_cap} for 1024-QAM. Such channel was chosen because we found that it emphasizes the advantage of the OFDM scheme for a certain range of SNR, even though its capacity is less than 1\% superior to the capacity of the ideal SC-DFE and it also generates a noticeable advantage for SC-DFE when higher SNR regimes are considered. In general, for 64- and 256-QAM, the OFDM capacity advantage is even smaller, when existent. In particular, for 64-QAM, the second derivative of $\tau(x)$ is positive for a small interval of $x$ comprised between 2.568 and 2.724, where it attains a maximum value of $3.58\times10^{-4}$. Thus, for practical considerations, the SC-DFE capacity can be considered equal or superior to the OFDM capacity for any channel for such modulation cardinality.

\section{Conclusions}
\label{sec:conclusions}

In this letter, we have compared the channel capacities of the ideal SC-DFE and OFDM schemes for square $M$-QAM modulations in frequency selective channels. Initially, we show that the subcarriers that are close to the capacity limit of the used modulation cannot fulfill the expected capacity to compensate for the attenuated subcarriers. Then, using Jensen's inequality and the square $M$-QAM capacity, we were able to prove that the ideal SC-DFE capacity is larger than the OFDM capacity for 4- and 16-QAM for any given channel. We have also analyzed higher-order square QAM and, in such a case, the OFDM scheme may surpass the ideal SC-DFE capacity, but only for a very small amount and only in specific scenarios. 


\bibliographystyle{IEEEbib}
\bibliography{bibli2}

\end{document}